# Reversible and magnetically unassisted voltage-driven switching of magnetization in FeRh/PMN-PT

*Ignasi Fina,*[1, a] *Alberto Quintana,*[2] *Xavier Martí,*[3] *Florencio Sánchez,*[1] *Michael Foerster,*[4] *Lucia Aballe,*[4] *Jordi Sort* [2,5] *and Josep Fontcuberta*[1]

[1]Institut de Ciència de Materials de Barcelona (ICMAB-CSIC), Campus UAB, E-08193 Bellaterra, Barcelona, Spain

[2]Departament de Física, Universitat Autònoma de Barcelona, E-08193 Bellaterra, Barcelona, Spain

[3]Institute of Physics, Academy of Sciences of the Czech Republic, Cukrovarnická 10, 162 53 Praha 6, Czech Republic

[4]ALBA Synchrotron Light Facility, Carrer de la Llum 2-26, Cerdanyola del Vallès, Barcelona 08290, Spain

[5]Institució Catalana de Recerca i Estudis Avançats (ICREA), Pg. Lluís Companys 23, E-08010 Barcelona, Spain

Reversible control of magnetization by electric fields without assistance from a subsidiary magnetic field or electric current, could help reduce the power consumption in spintronic devices. When increasing temperature above room temperature, FeRh displays an uncommon antiferromagnetic to ferromagnetic phase transition linked to a unit cell volume expansion. Thus, using the strain exerted by an adjacent piezoelectric layer, the relative amount of antiferromagnetic and ferromagnetic regions can be tuned by an electric field applied to the piezoelectric material. Indeed, large variations in the saturation magnetization have been observed when straining FeRh films grown on suitable piezoelectric substrates. In view of its applications, the variations in the remanent magnetization rather than those of the saturation magnetization are the most relevant. Here, we show that even in the absence of any bias external magnetic field, permanent and reversible magnetization changes as high as 34% can be induced by an electric field, which remain after this has been zeroed. Bulk and local magnetoelectric characterization reveal that the fundamental reason for the large magnetoelectric response observed at remanence is the expansion (rather than the nucleation) of new ferromagnetic nanoregions.

---

[a] ignasifinamartinez@gmail.com; ifina@icmab.es

To decrease the power consumption in spintronic memory devices is a fundamental requirement for coping with increased data storage capacity.[1,2] Among the different routes proposed for achieving this objective, perpendicular-to plane spin-transfer torque (STT) switching, is nowadays used as it allows lower energy dissipation and in-plane STT and more favorable scaling behavior than conventional Oersted fields[3]. On the other hand, multiferroic systems are interesting because they combine mutually coupled ferroelectric (FE) and ferromagnetic (FM) orders.[4-7] If coupling between the two orders exists, magnetization can be controlled without any electric charge flowing through the material, greatly reducing the power required (due to the absence of Joule heating power dissipation). However, widely investigated single-phase room-temperature multiferroic materials[8,9] are scarce and/or display a weak magnetoelectric coupling. In contrast, FM/FE hybrid structures are advantageous for applications, because heterostructures can be engineered to display large magnetoelectric coupling at room temperature. Depending on the nature of the FM and FE constituents, magnetoelectric coupling in FM/FE heterostructures can arise through various mechanisms and combinations of these, including: modulation of carrier density by electric field-effect;[10,11] modification of the magnetic anisotropy by changing the hierarchy of the electronic orbitals and their electronic filling;[12,13] modification of the structure-controlled magnetic exchange interactions;[14,15] and direct electric control of magnetic exchange interactions.[16-18] Voltage-controlled strain-mediated coupling can also produce changes in the magnetic anisotropy, coercivity, or saturation magnetization in a tunable and robust way. Overall, it is believed that electric-field control of magnetization would be energetically more efficient than current control as used nowadays.[19,20] Indeed, Joule heating associated to the large current required in SST devices[21-23] appears to be a serious bottleneck that electric-field-driven magnetoelectric devices could potentially overcome.[19,20,24]

Moreover, although strain can dictate the magnetic easy axis, and consequently the magnetization direction, in the absence of an external magnetic field, it cannot determine its orientation (vectorial direction). Thus, there are intrinsic difficulties in manipulating magnetization using an electric field without additional biasing magnetic field.[25-29] Local voltage-controlled magnetization switching was observed in nanometer regions of nickel layers embedded in a piezoelectric matrix due to the presence of internal magnetic fields.[30] More recently, predictions have been made suggesting that under restricted conditions, fast electric manipulation of magnetic moment orientation, remaining robust against thermal fluctuations, could be observed.[31-33] Nonetheless, experimentally demonstrating the reversible manipulation of bulk magnetization by electric fields with no auxiliary magnetic fields in strain-mediated systems remains a challenge.

A particular class of magnetoelectric system in which a large strain-mediated magnetoelectric coupling has been shown, is the FeRh/piezoelectric heterostructure. $\alpha'$-FeRh is an alloy that displays a first-order phase transition from antiferromagnetic (AFM) to FM order upon warming (at the Néel temperature, $T_N \approx 75°C$), and the concomitant coexistence of AFM and FM nanoregions and thermal hysteresis across the phase transition has been documented.[34] The unit cell volume expansion (about 1%)[35] associated with the phase transition is the lever that allows large strain-mediated magnetoelectric coupling. In this case, by transferring a suitable strain from a piezoelectric substrate to the FeRh film (by means of an electric field applied to the piezoelectric substrate), the relative amount of coexisting AFM-FM phases near $T_N$ can be tuned. Consistently, a modulation of saturation magnetization and electrical resistivity has been observed in FeRh films grown on top of piezoelectric substrates.[36-43] It is interesting to note that all the aforementioned studies exploited the piezoelectric character of the substrate, which was employed as a mere actuator. However, none of them exploited its FE and thus hysteretic piezoelectric character. This could actually open the door to non-volatile memory devices.[44]

In the race towards memory applications, it is essential to know whether the magnetic moment *at remanence* (when all external stimuli, either magnetic or electric, have been zeroed) can be set *ad hoc* by an electric field in a robust and predetermined manner. As mentioned, strain can only determine the direction of magnetization (i.e., in-plane *versus* out-of-plane or in-plane rotation by 90º) but not its orientation (left/right or up/down). For this reason, in the case of FeRh alloy, the magnetization direction of piezoelectrically generated FM domains should randomly point towards opposite orientations and thus, at remanence, the overall strain-induced magnetization should be zero. In a shocking contrast with the previous statement, here we show that tunable and non-volatile magnetic states in FeRh can be obtained at remanence. We show that the observed magnetoelectric effect and the observed memory effect are related to the expansion of the FM domains. Moreover, in some regions of the sample, magnetic imaging also revealed the 180º magnetization switch processes that we argue could be related to FE domain wall motion.

A 50 nm thick FeRh film was grown onto a 2 mm x 3 mm (0.72)[PbMg$_{1/3}$Nb$_{2/3}$O$_3$]-(0.28)[PbTiO$_3$] (PMN-PT) (001)-oriented single-crystal substrate (Atom Optics Co) 500 μm-thick, by sputtering at 375ºC in an Ar atmosphere (0.01 mbar), followed by *in-situ* annealing at 500 ºC and 0.1 mbar for 1 hour using 10 ºC/min heating and cooling ramps.[38,39] The relevant magnetic and elastic parameters for FeRh and PMN-PT are summarized in Supplementary Table S1. The net in-plane magnetization was measured as a function of temperature and applied voltage using a vibrating sample magnetometer (VSM) from MicroSense Co (LOT-Quantum Design). Electric voltage and surface charge were applied and measured using a 617 electrometer from Keithley Co. For electric biasing, the FeRh/PMN-PT sample was contacted using the FeRh film as top electrode, while the bottom of the PMN-PT was covered with silver paste as shown in Figure 1(a). X-ray magnetic circular dichroism in combination with photoemission electron microscopy (XMCD–PEEM) experiments were performed at the CIRCE beamline of the ALBA Synchrotron[45] using circularly polarized x-rays with an energy resolution of $E/\Delta E \approx 5000$ on a sample grown under the same conditions but additionally capped by a thin AlO$_x$ layer in order to minimize oxide formation on the FeRh surface when exposed to the air. All XMCD–PEEM images were recorded at the Fe L3 edge at ≈707 eV.

First, we isothermally measured the electric field dependence of the in-plane magnetic moment (M) of FeRh/PMN-PT using the VSM. Prior to the magnetoelectric measurement, the sample was heated to 180 °C and cooled down to the measuring temperature (110 °C), slightly above the AFM-FM phase transition of FeRh, under a 10 Oe magnetic field to set a defined initial magnetic state ($M = M_{ini}$). When the set temperature was reached, the external magnetic field was zeroed and the magnetic moment of the sample was measured as a function of the applied bias voltage ($V_{bias}$). A FE polarization versus applied voltage (*P-V*) loop previously recorded at this temperature [Figure 1(b)] is used as a reference for the subsequent experiments. Figure 1(b) shows that the polarization is around 30 µC/cm$^2$ and the coercive voltage ($V_C$) is around 60 V (1.2 kV/cm). $V_{bias}$ was applied following the sequence indicated in Figure 1(c), where the solid line denotes the applied voltage and spheres signal the instants where the magnetic measurements were performed. In Figure 1(d) the relative change of magnetization [$\Delta M(V_{bias})/M_{ini} = (M(V_{bias})−M_{ini})/M_{ini}$] is plotted as a function of $V_{bias}$. It can be seen that when increasing $V_{bias}$ (from -100 V to +100 V, solid squares) the magnetic moment remains constant up to about +50 V, where [$\Delta M(V_{bias})/M_{ini}$ largely increases, reaching a maximum at $V_{bias} \approx 60$ V and decreasing when $V_{bias}$ is increased further. Along the voltage return path (from +100 V to -100 V, open squares), the magnetic moment displays a symmetric behavior, reaching a maximum at $V_{bias} \approx -60$ V. The most remarkable result is that $\Delta M(V_{bias})/M_{ini}$ shows a finite variation with $V_{bias}$. As mentioned in the introduction, in a strain-mediated system and in absence of any biasing magnetic field, strain-induced FM domains should be formed with a random direction of magnetization and thus $M(V_{bias})/M_{ini}$ to be zero. This is contrary to the results presented in Figure 1(d). In addition, $\Delta M(V_{bias})/M_{ini}$ displays two maxima at ±60 V and the position of these maxima coincides with the coercive fields (±$V_C$) observed in the FE loop shown in Figure 1(b). This result strongly indicates that strain determines the magnetization of the sample (hence, the observed magnetoelectric coupling is indeed strain-mediated), without an appreciable presence of spurious effects.[14,46] Note that any interfacial oxidation/reduction processes that could contribute to the change of magnetization, would be an odd function of the applied voltage, or irreversible. Thus, magnetization would increase (or decrease) for positive voltage and decrease (or increase) for negative. Figure 1(d) shows that this clearly did not happen, indicating that if interfacial oxidation *in-operando* conditions occur, this is not relevant for the magnetization dependence on $V_{bias}$ or $V_{pol}$. Another interesting result is that when $V_{bias}$ differs significantly from $V_C$, $\Delta M(V_{bias})/M_{ini}$ is extremely small indicating the important role of FE domain walls, which are more abundant near $V_C$.

Next, we address the stability of the electrically written magnetic configuration by measuring the magnetization after applying a $V_{bias}$ and subsequently zeroing it. In this way, we defined poling voltage ($V_{pol}$) as the amplitude of the voltage pulse applied before measuring magnetization at 0 V. Accordingly, we repeated similar voltage-dependent experiment to that in Figure 1(c,d), but zeroing the bias voltage before each magnetization measurement using the sequence shown in Figure 1(e). The values of $\Delta M(V_{pol})/M_{ini}$, measured at 0 V, as a function of $V_{pol}$ are shown in Figure 1(f). We emphasize the fact these data were collected at electric and magnetic remanence, that is at $V = 0$ and $H = 0$. Figure 1(f) shows that $\Delta M(V_{pol})/M_{ini}$ displays peaks at similar voltages and having similar magnitudes to those obtained with the sample under the $V_{bias}$ applied [Figure 1(d)]. Therefore, a voltage applied on the PMN-PT substrate produces a reversible and permanent modulation of the magnetization without applying an external magnetic field. We tested the retention of the state established by $V_{pol}$ and it was found that it lasts at least $2 \times 10^3$ s (see Supplementary Figure S1).

We recall here that at $V_C$, the electrically induced in-plane deformation of the substrate is minimal, whereas at $V >> |\pm V_C|$ it is at its maximum compression. Consequently, at $V_C$ the FM phase of FeRh should be more abundant, whereas the higher compressive strain for $V >> \pm V_C$ favors the AFM phase [Figure 2(a)]. It therefore follows that $\Delta M(V_{pol})$ should have maxima at $V = \pm V_C$, as previously shown in Figure 1(d,f). It is important to notice that the observed $\Delta M(V_C)$ maxima indicate a increase of magnetization, thus implying that the strain-induced FM regions have the magnetization pointing along that of preexisting FM domains. Further evidence of this will be shown latter. Probably, the most clear evidence of the capability of the non-volatile writing of different magnetic states by using a FE substrate, which is not attainable with a piezoelectric/paraelectric actuator, is to perform minor $\Delta M(V)$ hysteresis loops. The $\Delta M(V_{pol})/M_{ini}$ minor loop, obtained by a $V_{pol}$ excursion from -100 V to +60 V and back to -100 V, allows two different strain states to be reached at electric remanence, as well as a magnetization modulation of around 30% as shown in Figure 2(b). The energy invested to complete the $\Delta M(V_{pol})/M_{ini}$ loop is that corresponding to the energy invested to produce a minor FE loop, which corresponds to $\sim 72$ kJ/m$^3$ (see Supplementary Appendix S1 for a detailed explanation of the energy evaluation). This value is slightly larger than that reported for similar strain mediated systems.[24]

We next focus on the modulation of the magnetoelectric coupling and its efficiency, and we show that the change of magnetic moment $\Delta M$ achievable at a given $V_{pol}$ is determined by the initial amount of FM phase preexisting in the sample. Distinct initial $M_{ini}$ can be obtained by varying the magnetic field during the cooling process prior to the measurement. The dependence of $\Delta M(M_{ini})$ on the initial magnetic state is summarized in Figure 3(a). $V_{pol} = 60$ V is the voltage at which the magnetoelectric coupling and thus $\Delta M(V_{pol})$ are at their maxima. It is observed that $\Delta M$ increases with $M_{ini}$ and eventually saturates (at $M_{ini} \approx 728$ emu/cm$^3$). It is interesting to observe that although $\Delta M$ depends on $M_{ini}$, its relative variation ($\Delta M/M_{ini}$) is somewhat constant [circles in Figure 3(b)] and $\Delta M/M_{ini}$ reaches an average value of 34% almost independently of $M_{ini}$. Figure 3(a) includes similar data obtained using smaller (50 V) and larger (70 V) pulses and the $\Delta M$ changes are found to be smaller (in agreement with $V_C \approx 60$ V). To address the efficiency of the magnetoelectric effect, Figure 3(c) shows the $M(H)$ loops recorded at -100 and 60 V. It is worth noting that the variation of the remanent magnetic moment after saturation is $\Delta M_r \approx 100$ emu/cm$^3$. This value is similar to the maximum increase of $\Delta M \approx 100$ emu/cm$^3$ in Figure 3(a) (obtained with $M_{ini} = 356$ and 728 emu/cm$^3$), indicating that the magnetic moment of all the new FM areas produced by $V_{pol}$ point in the same direction as the preexisting moment if it is high enough.

The observation that $\Delta M$ is proportional in magnitude and orientation to the preexisting $M_{ini}$ [Figure 3(a)] indicates that the magnetization of the *new* strain-induced FM domains follows that of the pre-existing FM domains, suggesting that the new electrically-induced FM domains are formed adjacent to them. This infers that $\Delta M$ is governed by strain-controlled FM domain expansion, where old FM regions determine the magnetization orientation of the new ones, rather than random domain nucleation. If expansion of FM domains dominates over nucleation in the increase of $\Delta M$ with $V_{pol}$, one could anticipate that at lower temperatures, as the number of initial FM domains is smaller, the $\Delta M$ should be smaller too, because random nucleation is more important and not contributing to $\Delta M$. To verify this prediction, we performed $\Delta M(V_{pol})$ measurements at a lower temperature (75 °C). As shown by the data [see Supplementary Figure S2(b)], a smaller variation $\Delta M(V_{pol})$ was indeed found, although the variation of the saturation magnetization with electric field was near its maximum [see Supplementary Figure S2(a)]. We therefore conclude that the expansion of FM domains by voltage-controlled strain accounts for the observed large magnetoelectric response at remanence at high temperatures (110 °C).

To reinforce our interpretation of the observed magnetoelectric effects, nanoscale magnetic domain arrangements were imaged through XMCD-PEEM experiments 110 °C, with the sample under $V_{bias}$ with the aim of examining the influence of applied voltage on the magnetic domain motion and distribution. A representative XMCD-PEEM image collected at $V_{bias}$ = -100 V is shown in Figure 4(a), where red and blue regions illustrate the FM in-plane domains pointing towards the left and right orientations, respectively. White regions account for ferromagnetic domains having its magnetization axis perpendicular to the incident beam (up/down) or antiferromagnetic domains. Sequentially recorded images at $V_{bias}$ = -100, 10, 20, 40, 60, -20, -60, and -80V, respectively are shown in Supplementary Video 1. In Figure 4(b), the XMCD contrast ($XMCD = (I-I_{MAX})/I_{MAX}$ is plotted as a function of $V_{bias}$, where $I$ and $I_{MAX}$ are the average intensity and the maximum intensity in the region of interest (ROI), respectively, representing either predominantly red/white or blue/white [indicated in Figure 4(a)]. The intensity of the red/white region increases as $V_{pol}$ increases up to +60 V, meaning the red region expands. The intensity of the blue/white region decreases as $V_{bias}$ increases up to +60 V, meaning the blue region expands. Therefore, FM regions increase at the expense of AFM when increasing $V_{bias}$ up to +60 V in agreement with Figure 2(a). It can be seen that both regions describe a hysterical curve, which locally mimics that shown in Figure 2(b). These contraction/expansion effects are similar to those reported in pure magnetic systems, where the magnetic dynamics are triggered by the magnetic field.[47,48]

Finally, we note that in some limited regions of the FeRh film, the XMCD-PEEM images display a remarkably different magnetic behavior. Indeed, upon V cycling, a switch by 180º of the magnetization is observed. As example, in Figure 4(c), we show the XMCD-PEEM image collected at $V_{bias}$ = -100 V on one of those regions. The area delineated by the inclined dashed lines is a zone were most of magnetization reverses its sign (contrast change) upon V cycling. This can be better appreciated by analyzing the XMCD contrasts change in a limited zone in this region (for instance the square area indicated) and plotting the magnetic contrast upon $V_{bias}$ cycling. This is shown in Figure 4(d), were we plot the XMCD normalized intensity as a function of $V_{bias}$. Sequential XMCD images recorded at $V_{bias}$ = -100, 10, 20, 40, 60, -20, -60, and -80V, respectively, shown in Supplementary Video 2, more clearly show the magnetization switching in this regions. A hint to the microscopic origin of the distinct magnetic response of these regions, that is reversion of magnetic moment rather that expansion/contraction of FM domains as seen in Figure 1 and Supplementary Information 1, can be obtained by the topographic images recorded at $V_{bias}$ = -100, 10, 20, 40, 60, -20, -60, and -80V, respectively shown in Supplementary Video 3. Detailed inspection of these images suggest that a domain wall propagation occurs in these regions and thus important strain gradients shall be present.

In conclusion, we have demonstrated that by exploiting the piezoelectric and FE character of the PMN-PT, the macroscopic magnetic moment of a FeRh film can be permanently modulated at electric and magnetic remanence by about 34%. Using suitable electric and magnetic conditioning, the magnetic state of the sample can be brought to different remanent states, and both the direction and orientation (and hence the magnitude) of the remnant magnetization can be tailored at will. Magnetic characterization and imaging has shown that the electrically stimulated expansion/contraction of FM domains is the driving mechanism for the observed permanent magnetization modulation.

## Supplementary material

Supplementary Figure S1 includes magnetization retention measurements. Supplementary Figure S2 includes complementary magnetoelectric characterization. Supplementary Appendix 1 includes a description of the calculation of the energy invested in the minor loop in Figure 2b. Supplementary Table 1 summarizes the significant parameters of the studied system.

Supplementary Video 1 shows the evolution of magnetic domains in the region shown in Figure 4(a). (top panel) XMCD-PEEM images collected at -100, 10, 20, 40, 60, -20, -60, and -80V in the same region as that in Figure 4(a). (Middle panel) Evolution of the $V_{bias}$ with time. Each frame indicates the $V_{bias}$ applied. (Bottom panel) The evolution of the intensity normalized to its maximum value for each frame for the indicated regions BLUE for predominant blue/white, and red for predominant red/white. In the video the sequence is repeated (with the same data) for increasing and decreasing time three times to better visualize time evolution. It shows the expansion of both the red and blue regions at high positive $V_{bias}$.

Supplementary Video 2 shows the evolution of the magnetic domains in the region shown in Figure 4(c). (Top panel) XMCD-PEEM images collected at -100, 10, 20, 40, 60, -20, -60, and -80V in the same region as that in Figure 4(a). (Middle panel) Evolution of the $V_{bias}$ with time. Each frame indicates the $V_{bias}$ applied. (Bottom panel) Evolution of the intensity normalized to its maximum value for the indicated region. The area delineated by the inclined dashed lines is a zone were most of magnetization reverses its sign. The yellow arrow indicates the presence of topographic deformation and its evolution inferred from Supplementary Video 3. In the video the sequence is repeated (with the same data) for increasing and decreasing time three times to better visualize time evolution. It shows the inversion of the blue region (at high negative $V_{bias}$) into red (at high positive $V_{bias}$).

Supplementary Video 3 shows the topographic evolution of the same region as that shown in Figure 4(c). (Top panel) Topographic PEEM images collected at -100, 10, 20, 40, 60, -20, -60, and -80V in the same region as that in Figure 4(a). The yellow arrow indicates the presence of topographic deformation and its evolution. The area delineated by the inclined dashed lines is a zone were most of magnetization reverses its sign. (Middle panel) Evolution of the $V_{bias}$ with time. Each frame indicates the $V_{bias}$ applied.

# Acknowledgements

Financial support by the Spanish Government [Projects MAT2014-56063-C2-1-R, MAT2015-73839-JIN, MAT2014-51778-C2-1-R, MAT2015-64110-C2-2-R and MAT2017-86357-C3-1-R, and associated FEDER], the Generalitat de Catalunya (2017-SGR-1377, 2017-SGR-292), and the European Research Council (SPIN-PORICS 2014-Consolidator Grant, Agreement nº 648454) is acknowledged. We also acknowledge support from EU ERC Advanced Grant No. 268066, from the Ministry of Education of the Czech Republic Grant No. LM2011026, from the Grant Agency of the Czech Republic Grant no. 14-37427. ICMAB-CSIC authors acknowledge financial support from the Spanish Ministry of Economy and Competitiveness, through the "Severo Ochoa" Program for Centers of Excellence in R&D (SEV- 2015-0496). I.F. acknowledges Ramon y Cajal contract RYC-2017-22531.

# Figure 1

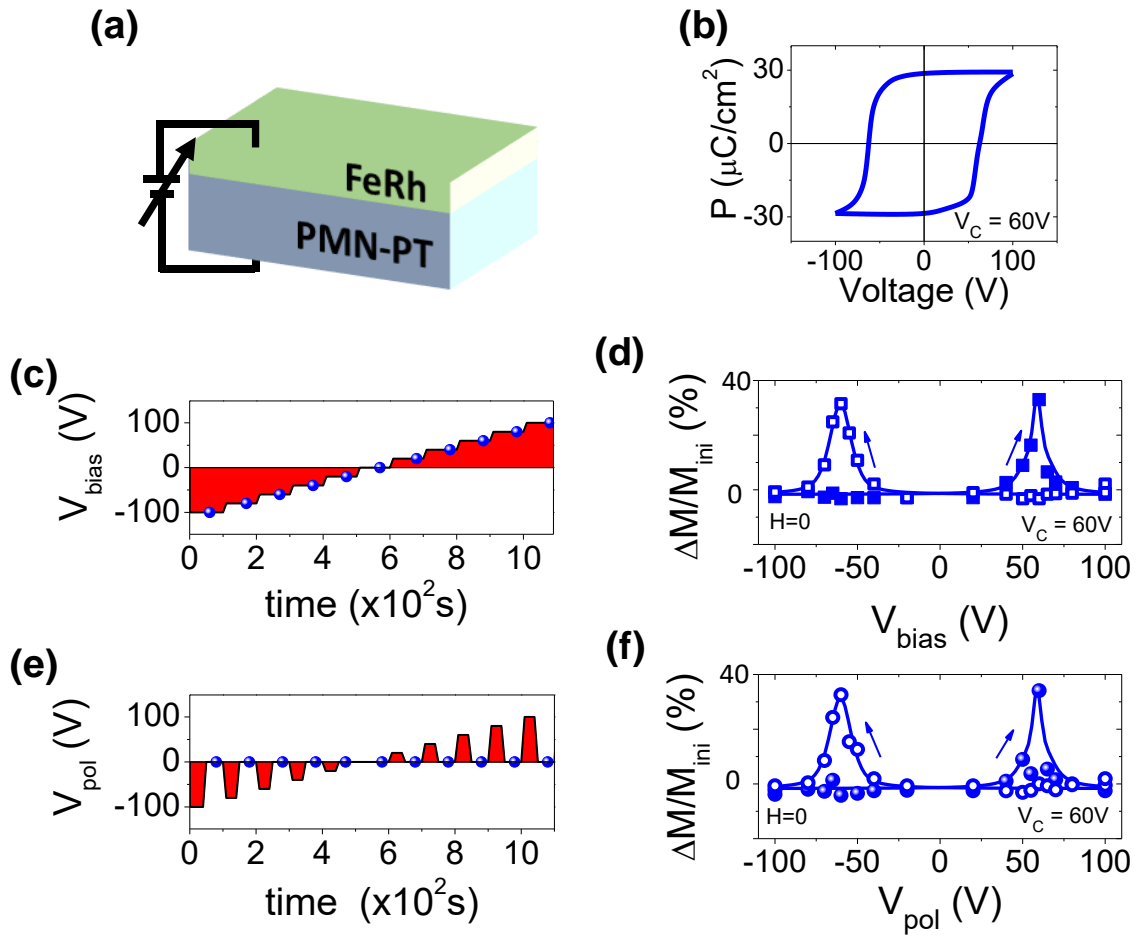

*Figure 1.* (a) Sketch of the electric connection to the FeRh film on PMN-PT(001) to apply the external bias voltage. (b) Polarization versus voltage loop at 110ºC. (c) Applied voltage dependence on time (for increasing voltage) for the data displayed in (d), blue spheres correspond to the magnetic moment measurement. (d) Relative magnetic moment increase with respect to the magnetic moment in the initial state [$\Delta M/M_{ini} = (M-M_{ini})/M_{ini}$] depending on bias voltage ($V_{bias}$), solid squares for increasing bias voltage and empty squares for decreasing bias voltage. (e) Applied voltage dependence on time (for increasing voltage) for the data displayed in (f), blue spheres correspond to the magnetic moment measurement at zero $V_{bias}$. (f) Relative magnetic moment increase with respect to the magnetic moment in the initial state at electric remanence (0 V) depending on previously applied bias voltage ($V_{pol}$), solid circles for increasing bias voltage and empty circles for decreasing bias voltage.

**Figure 2**

**(a)**

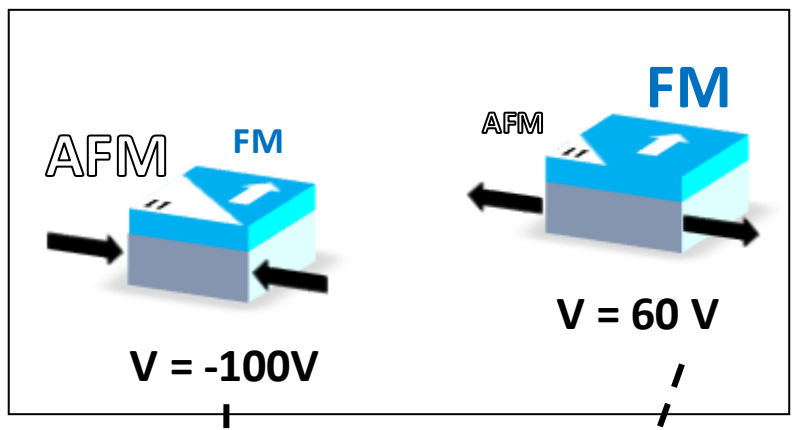

**(b)**

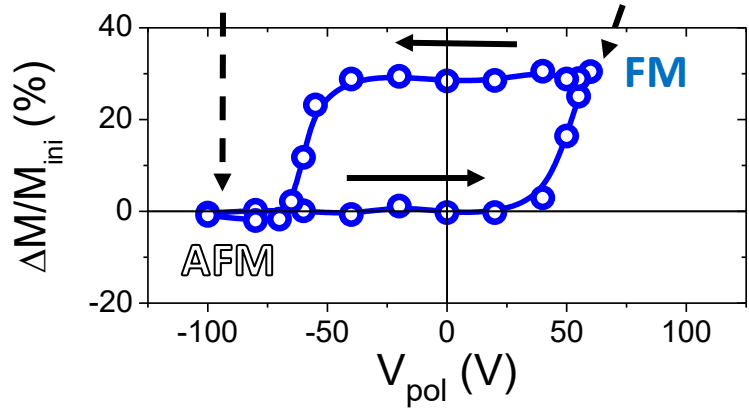

*Figure 2. (a) Sketch of the two remnant magnetization states obtained for the two different strain states. The increase/decrease of the substrate in-plane strain induced by the application of the external voltage (poled or $V_{bias} = V_C = +60$ V), and the concomitant FeRh unit cell expansion/contraction is sketched. Blue and white indicate FM and AFM regions, respectively. (b) Relative magnetic moment increase (with respect to the initial magnetic moment voltages) for several applied voltages from -100 to 60 and back to -100V measured following the arrows direction.*

**Figure 3**

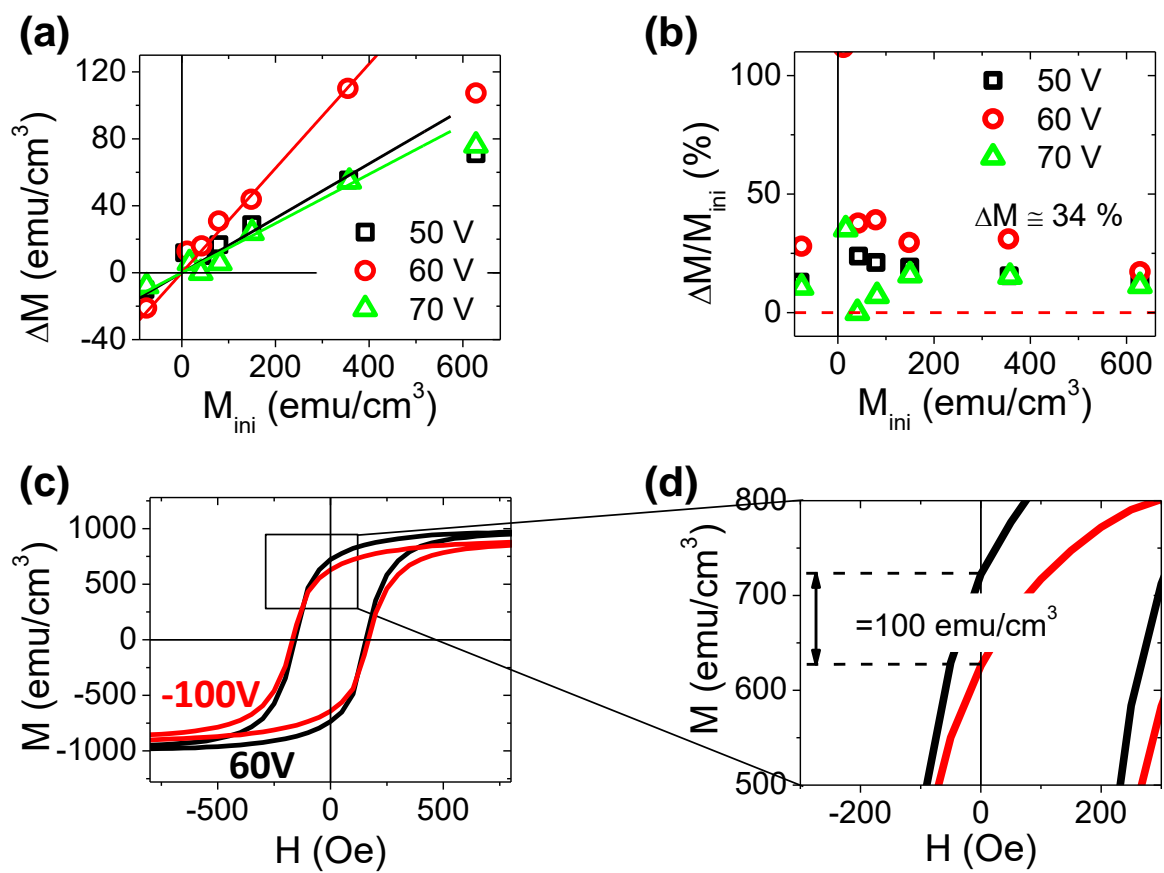

*Figure 3.* (a) Dependence of ΔM on $M_{ini}$ for V = 50, 60 and 70 V as indicated in the legend. (b) Dependence of relative ΔM/$M_{ini}$ on $M_{ini}$ for V = 50, 60 and 70 V as indicated in the legend. (c) M(H) loops for -100 V and +60 V (applied after poling the sample with -100V at 110 °C). (e) Zoom of (d) showing that the variation of remanent magnetic moment after saturation by electric field is 100 emu/cm$^3$, as it is the maximum magnetic moment variation found in (a).

**Figure 4**

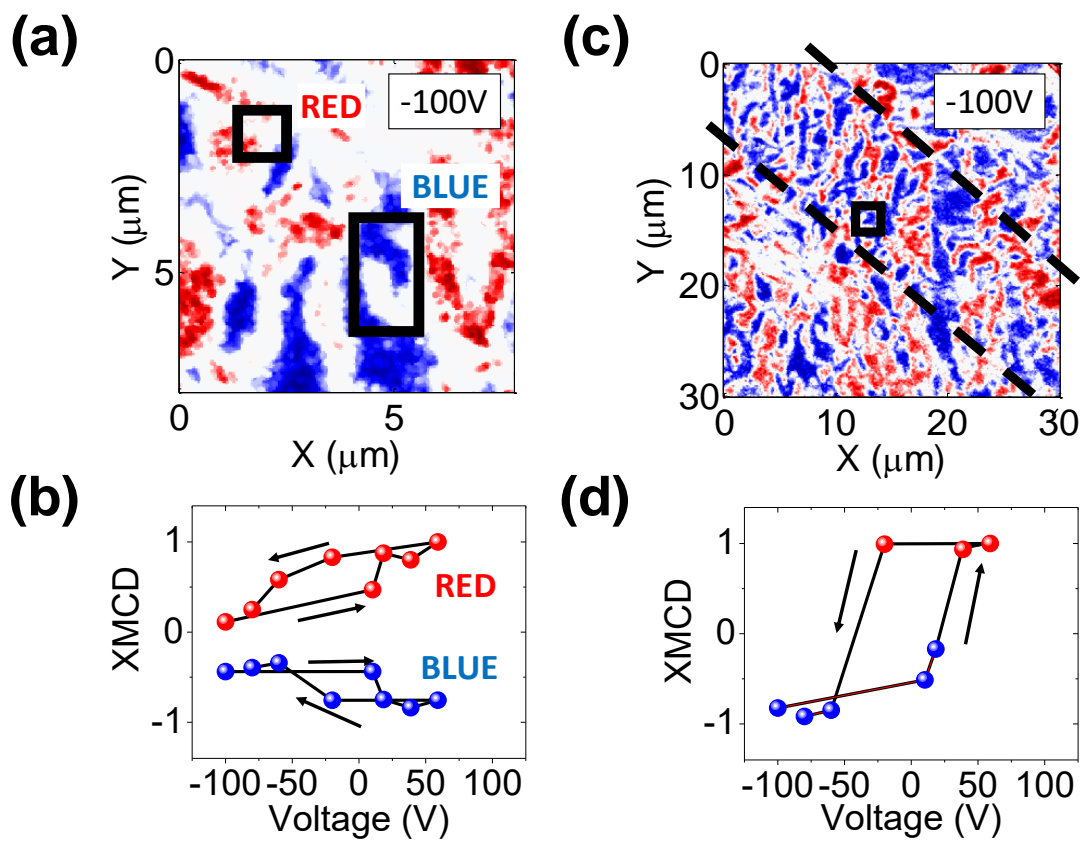

*Figure 4.* (a) XMCD-PEEM images at $V_{bias}$ = -100V for a representative region of the sample. (b) XMCD (XMCD = $(I-I_{MAX})/I_{MAX}$) contrast of the signaled regions in (a). (c) XMCD-PEEM images at $V_{bias}$ = -100V for a region of the sample where 180º switching was detected. (d) XMCD (XMCD = $(I-I_{MAX})/I_{MAX}$) contrast of the signaled region in (c). Enclosed regions are always between 1 and 2 $\mu m^2$ for better comparison.

# Supplementary material of:

# Reversible and magnetically unassisted voltage-driven switching of the magnetization in FeRh/PMN-PT

*Ignasi Fina,*[1, b] *Alberto Quintana,*[2] *Xavier Martí,*[3] *Florencio Sánchez,*[1] *Michael Foerster,*[4] *Lucia Aballe,*[4] *Jordi Sort*[2,5] *and Josep Fontcuberta*[1]

[1]Institut de Ciència de Materials de Barcelona (ICMAB-CSIC), Campus UAB, E-08193 Bellaterra, Barcelona, Spain

[2]Departament de Física, Universitat Autònoma de Barcelona, E-08193 Bellaterra, Barcelona, Spain

[3]Institute of Physics, Academy of Sciences of the Czech Republic, Cukrovarnická 10, 162 53 Praha 6, Czech Republic

[4]ALBA Synchrotron Light Facility, Carrer de la Llum 2-26, Cerdanyola del Vallès, Barcelona 08290, Spain

[5]Institució Catalana de Recerca i Estudis Avançats (ICREA), Pg. Lluís Companys 23, E-08010 Barcelona, Spain

---

[b] **ignasifinamartinez@gmail.com**

**Figure S1.** ΔM/M$_{ini}$ value read after indicated delay time after V$_{pol}$ application. At 0 delay time, the initial state value is included.

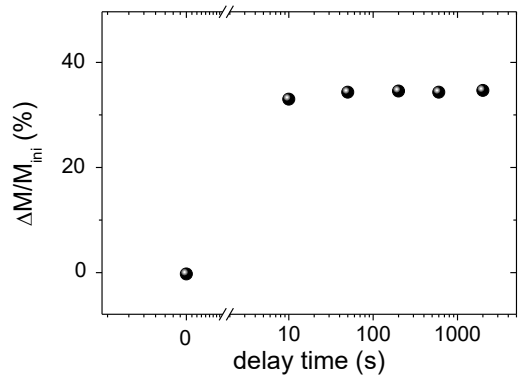

**Figure S2.** (a) Absolut difference of the saturated magnetization (ΔM$_{sat}$) curves recorded at 1000 Oe (safely above H$_C$) under 0 and 100 V. It can be observed that a minimum (maximum magnetoelectric effect) is reached at near 80 °C. Solid lines indicate the 75 °C at which data shown in (b) is collected. Dashed line indicates 110 °C at which the measurements displayed in the main text are performed. It can be inferred that the maximum ΔM$_{sat}$ is only two times that shown in the experiment performed at remanence at 110 °C. (b) Comparison of the relative magnetic moment increase for measurements performed at 110 °C and various M$_{ini}$, and at 75°C for different voltage pulses. It can be observed that at 75°C the effect is always lower.

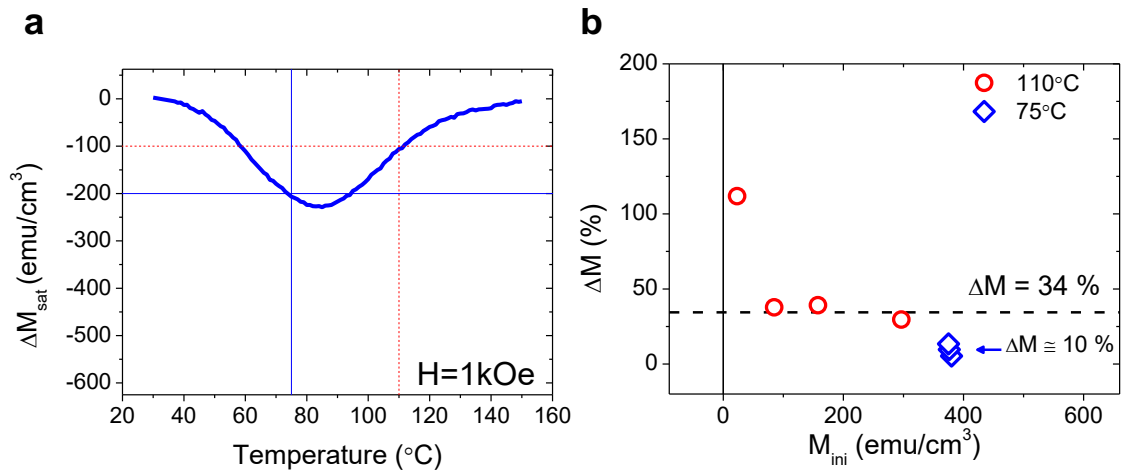

**Table S1.** Summary of geometrical dimensions and relevant parameters (piezoelectric coefficients ($d_{ij}$) and Young modulus) of FeRh in thin film form and PMN-PT single-crystal with (001) orientation. The given parameters of magnetization ($M_S$) and in-plane magnetic anisotropy constant ($K_u$) of FeRh, correspond to those of the ferromagnetic (high temperature) phase. Further details can be found in the indicated references. $\alpha$ accounts for the damping coefficient. $Y$ is the Young modulus

| PMN-PT single crystal | | | FeRh film | | |
|---|---|---|---|---|---|
| Parameter | Value | | Parameter | Value | |
| *thickness* | 500 μm | | *thickness* | 50 nm | |
| *area* | 2 by 3 mm | | *area* | 2 by 3 mm | |
| $d_{33}$ | 1766 pC/N | 1 | $M_s$ | 1120 emu/cm$^3$ | 2 |
| $d_{31}$ | 723 pC/N | 1 | $K_u$ | -0.36 meV/cell | 3 |
| $d_{31}$ | -1761 pC/N | 1 | $\alpha$ | 0.02-0.03 | 4 |
| $Y$ | 18-20 GPa | 5,6 | | | |

**Appendix S1.** In the minor M-V loop displayed in Figure 2b, the input energy is the one necessary to produce the switching of the ferroelectric polarization. In our case ≈ 50% of P (because we are performing a minor loop) is switched (15 μC/cm$^2$), the coercive voltage is 60V, and the ferroelectric thickness 500 μm. Therefore, the electrostatic energy invested in this switching process and to produce the obtained magnetization variation is E=4·P·E$_c$=15 μC/cm$^2$·60V/500 μm = 72 kJ/m$^3$. In a 50 per 50 nm$^2$ cell with a thickness equal to the one of our sample (500 μm) the energy consumption would be 22 fJ. However, our device thickness can be in principle easily reduced to 500 nm in the worst case, which corresponds to 0.022fJ per unit cell. State of the art Spin Torque Transfer Magnetic memory with 5kΩ needs approx. 10 μA,[7] which is 0.5 fJ for each cell.

**References of supplementary material**